\documentclass{article}
\usepackage[english]{babel}

\usepackage{comment}

\usepackage[bibstyle=ieee,citestyle=numeric-comp,isbn=false,url=false,doi=false,eprint=false]{biblatex}
\usepackage[paper=a4paper,margin=0.8in]{geometry}

\usepackage[T1]{fontenc}
\usepackage[latin9]{inputenc}

\usepackage{amsbsy}
\usepackage{amstext}
\usepackage{amssymb}
\usepackage{authblk}
\usepackage{blindtext}
\usepackage{graphicx}
\usepackage{esint}
\usepackage{babel}

\providecommand{\tabularnewline}{\\}

\begin{document}

\date{}
\title{Markov State Models of protein-protein encounters}
\author[1]{Simon Olsson}

\affil[1]{Chalmers University of Technology, Department of Computer Science
and Engineering, R\"annv\"agen 6, 41258 G\"oteborg, Sweden.}
\affil[ ]{E-mail: simonols@chalmers.se}

\maketitle

The encounter of proteins is critical to countless biological processes
and may span several length- and time-scales \cite{Nooren_2003}.
For example, Insulin binds the $\alpha$-subunits of insulin receptors,
thereby activating the tyrosine-kinase $\beta$-subunit auto-phosphorylation
triggering a signal transduction cascade, leading to a broad range
of responses from the molecular, over cellular, and to the physiological
scales \cite{Czech1985}. Every step along this cascade involves protein-protein
interactions between different proteins or multiple copies of the
same protein chain. This example is just one of many illustrating
why mapping out the details of protein-protein encounters at the atomistic
and molecular scale is critical to understanding these processes,
what goes wrong in disease states, and inform intervention strategies
to remedy or reverse pathological conditions \cite{Zinzalla2009}.

Indeed, massive-scale efforts have attempted to characterize protein-protein
interaction networks using high-throughput experimentation \cite{Blikstad_2015},
and insights gained from these endeavors have undoubtedly been incredibly
impactful \cite{Gavin_2006,Krogan_2006}. However, these proteomic
approaches' strengths lie in their broad scope but not in their resolution.
Currently, only biophysical and molecular simulation techniques allow
us to dissect the intimate structural, thermodynamic, and kinetic
details \cite{Chakrabarti_2016,Weikl_2014,Palmer_2004,Min_2005,Frank_2018,Boehr_2009}.
For example, Cryogenic Electron Microscopy and X-ray crystallograph
may potentially give us high-resolution snapshots of the encounter
process at various stages \cite{Frank_2018,Dementiev_2006,Huntington_2000}.
Single-molecule FRET spectroscopy can give structural and kinetic
insights into protein-ligand binding \cite{Kim_2013}. Finally, Nuclear
magnetic resonance (NMR) spectroscopy enables detailed characterizations
of protein-protein encounters, possibly giving us structural, thermodynamic,
and kinetic insights, given favorable experimental conditions \cite{Sugase_2007,Chakrabarti_2016,Bezsonova_2008,Purslow_2020}.
Molecular dynamics simulations with explicit solvation uniquely give
us a fully spatiotemporally resolved view of protein dynamics \cite{Raich_2021,No__2009,Barros_2021,Lindorff_Larsen_2011,Shaw_2010,Stelzl_2020,Pietrucci2009,Stanley_2014}
including the encounter mechanism \cite{Plattner_2017,Pan_2019,Paul_2017}.
Advances in software and hardware technology enable us to routinely
reach aggregate simulation time-scales which overlap with experimental
time-scales for small protein-protein systems, especially when using
kinetic modeling approaches such as Markov state models (MSM) \cite{Prinz_2011,Husic_2018,Wehmeyer_2019}.

This chapter will outline how molecular dynamics simulations, experimental
data, and MSMs can synergize to map-out the mechanism of protein-protein
association and dissociation. Further, I will discuss whether we can
currently estimate accurate rates and thermodynamics of critical metastable
states. First, I motivate MSMs in the light of molecular dynamics
theory. Then I outline the practical aspects of applying MSMs to studying
protein-protein encounters and show some successful examples from
the literature. I will further discuss how to use experimental data
to validate and augment MSMs estimated from molecular simulation data.
I will close with a few examples of emerging technologies that may
improve computational study of protein-protein encounters in the future.

\subsection*{Notation}

\begin{tabular}{|l|l|}
\hline 
Symbol & Explanation\tabularnewline
\hline 
\hline 
$\mathbb{P}[x]$ & the probability of an event $x$ \tabularnewline
\hline 
$p(x)$ & a probability density function\tabularnewline
\hline 
$x\mid y$ & event $x$ given $y$. This may occur in probability densities or
probabilities of events\tabularnewline
\hline 
$\mathcal{P}_{\tau}$ & Markov Propagator. A 'continuous space equivalent' of a Markov state
model.\tabularnewline
\hline 
$\langle\rangle$ & Ensemble average with respect to the stationary distribution (Boltzmann
distribution)\tabularnewline
\hline 
$\mu(x)$ & Boltzmann distribution\tabularnewline
\hline 
\end{tabular}

\pagebreak{}

\section{Molecular dynamics and Markov state models }

When applying molecular dynamics simulations, we aim to understand
biomolecular processes. Ideally, our understanding must build on statistically
robust scientific observations. The key observables of interest:
\begin{enumerate}
\item Important structures,
\item their thermodynamic weights, 
\item and the transition probabilities amongst them, or their inter-conversion
rates.
\end{enumerate}
Robust identification of these three properties allows for MD results'
direct connection to experimental data, including NMR spectroscopy
and sm-FRET \cite{Olsson_2016,Noe_2011,Prinz_2011b,No__2008}. Comparisons
such as these may serve as an important complementary means of validating
the simulation models and can help drive robust scientific hypotheses
and models.

Analysis of MD simulations, however, often relies on visually inspecting
simulation trajectories one-by-one. Alternatively, we follow the simulation
trajectories projected onto a few order parameters (or collective
variables) derived from chemical intuition about the process of interest
or some global structural property \cite{Huber_1994,Grubm_ller_1995,Laio_2002,Rohrdanz_2013,Lange_2006}.
Inspecting structures and following certain order parameters is an
integral part of any analysis of molecular dynamics simulations. However,
these strategies alone do not guarantee a statistical relevance of
events observed, and the overall approach becomes increasingly time-consuming
with growing data-sets. Furthermore, limiting ourselves to these analyses
may still overlook rare events important for biological function.
So ultimately, conclusions drawn from these kinds of analyses may
be misleading \cite{Prinz_2011}.

Statistical models to analyze data from MD simulations are enjoying
increased attention in recent years \cite{Sch_tte_1999,Sch_tte_2011,Bowman2014,Buchete_2008,Swope_2004,Doerr_2016,Sriraman_2005,Zwanzig_1983,Rao_2004}.
This popularity is a necessary consequence of growing datasets enabled
by improvements in software efficiency and large-scale investment
into consumer-grade GPU (graphical processing units) based compute
resources by many academic groups. Another important factor is community-driven,
cloud-based super-computers such as Folding@Home \cite{Shirts_2000}
and GPUgrid (www.gpugrid.net) that generate enormous volumes of simulation
data whose analysis critically relies on a systematic and principled
framework. Markov state models (MSM) are one prominent example of
statistical models for analyzing molecular dynamics simulation, which
fits the bill \cite{Sch_tte_1999,Bowman2014,Prinz_2011,Chodera_2007}.

This section will briefly discuss the motivation and theoretical basis
of MSMs and some important mathematical properties of MSM, motivating
subsequent sections. With this text, I do not attempt to discuss these
topics comprehensively but instead, provide a guiding primer into
the following sections and enable the reader to build some intuition
about the theory -- in general, the text is based upon the references
cited in this section. However, I intentionally minimize technical
language and equations and avoid specific details in the notation
for clarity. For a more detailed MSM theory treatment, I refer to
the excellent review by Prinz et al. \cite{Prinz_2011}. For a more
comprehensive historical overview of MSMs, I refer to Brooke and Pande's
review \cite{Husic_2018}. A recent tutorial for step-by-step MSM
building is also available \cite{Wehmeyer_2019}.

\subsection{Markov state models: theory and properties}

Above, I outlined how we need to minimize the subjectivity going into
analyzing data from MD simulations. Such subjectivity may stifle our
ability to detect transient intermediate, or off-pathway, states,
parallel protein-protein association pathways, and other intricate
kinetic features. Consequently, we need simplification of the conformational
space needs to enable human interpretation of the results. However,
we should achieve this in a manner that supports our goals to extract
as much kinetic and thermodynamic information from our simulation
data as possible. MSMs provide a framework for achieving this goal.

But what \emph{is} a MSM? -- A MSM is an $N\times N$ matrix where
each element encodes the conditional probability of ending in a state
$i$ from state $j$ after a constant time, $\tau$ \cite{Sch_tte_1999}.
The $N$ states each represent a different disjoint segment of the
configurational space. Therefore, the MSM give us an \emph{ensemble}
\emph{view} of the molecular dynamics, where each trajectory corresponds
to a sample from a distribution of dynamics trajectories \cite{Prinz_2011}.
This view is exactly analogous to that taken in statistical mechanics
and thermodynamics: the accuracy at which we can characterize the
molecular system's important properties is limited by how well we
can estimate the statistical distribution of the system's dynamics.
Any given trajectory will typically be too short to be representative
of the full system dynamics. However, estimating this statistical
distribution would allow us to pinpoint important structures and compute
their thermodynamic and kinetic properties. The statistical distribution
further allows us to predict how a non-equilibrium initial condition
--- prepared in an experiment --- relaxes back to equilibrium or
predicts experimentally measurable spectroscopic observables\cite{Prinz_2011b,Olsson_2016,Noe_2011}.

At first glance, estimating this statistical distribution may seem
completely infeasible: the distribution domain is all possible temporal
trajectories of a molecular system with all-atom detail. To make this
estimation tractable, we rely on the following assumptions:
\begin{enumerate}
\item Time-homogenous Markovian dynamics
\item Ergodicity
\item Reversibility.
\end{enumerate}
The first assumption restricts the dynamics we can consider to one
where the transition probabilities from $x_{t_{M}-t}$ to $x_{t_{M}}$
after some time $t$ are independent of what happened before. These
transition probabilities do not change with time. More formally, we
can simplify the conditional probability of arriving in $x_{t_{N}}$
given all prior states, $x_{t_{0}},\dots,x_{t_{M}-t}$, by,
\[
\mathbb{P}[x_{t_{N}}\mid x_{t_{0}},\dots,x_{t_{M}-t},t]=\mathbb{P}[x_{t_{M}}\mid x_{t_{M}-t},t]
\]
that is, the probability of arriving in a state at time $t_{N}$ only
depends on the state the system was in at $t_{M}-t$ and that this
probability is \emph{invariant} to a time-shift -- \emph{homogeneous}.
A trajectory of a systems dynamics is here represented by the states
the system adopts $x_{t_{0}},\dots,x_{t_{M}-t}$ at a sequence of
sampled uniformly in time $t_{0},\dots t_{M}-t$.

The second assumption tells us that we can reach any point in configuration
space from any other point configuration space within some finite
time. There is a non-vanishing probability of arriving at any state
$x'$ from any other state $x$ in a finite time. This assumption
ensures the configuration space to be \emph{dynamically connected}.

The final assumption ensures that the probability flux between points
$x$ and $x'$ in configuration space is the same in either direction.
In physical terms, this means that energy is not extracted or generated
in any state. Formally, this corresponds to the fulfillment of the
\emph{detailed balance }condition

\[
\mathbb{P}[x']\mathbb{P}[x\mid x']=\mathbb{P}[x]\mathbb{P}[x'\mid x]
\]
where $\mathbb{P}[x]$ is the stationary --- equilibrium -- probability
of state $x$, typically given by the Boltzmann probability $\mathcal{Z}^{-1}\exp(-\beta U(x))$
for molecular systems at thermal equilibrium. I have suppressed the
time dependence of the conditional transition probability for notational
brevity. Strictly, this final assumption is unnecessary as many simulation
setups involve doing work on the molecular system. In such scenarios,
other factors will drive the system beyond the thermal fluctuations,
and, in general, the system will not be in thermal equilibrium. Nevertheless,
the fulfillment of the detailed balance condition leads to the symmetry
of the joint probability $\mathbb{P}[x,x']=\mathbb{P}[x',x]$, which
we will see allows for a more statistically efficient estimation of
MSMs in many cases.

Are all of these assumptions fulfilled in any practical cases? --
Yes! Most of the common thermostatting algorithms used are consistent
with the assumptions I outline above in molecular simulations. Prinz
et al. discuss notable exceptions \cite{Prinz_2011}.

Remember, our original goal was to arrive at an \emph{ensemble view}
of molecular dynamics. This view describes the time-evolution of many
copies of the same molecular system. The copies are independent, and
do not interact with each other, and are distributed according to
the Boltzmann distribution, when at equilibrium, $\mu(x)=\mathcal{Z}^{-1}\exp(-\beta U(x))$.
$\mathcal{Z}$ is the partition coefficient, $U(\cdot)$ is the system
potential energy at the experimental conditions, and is the inverse
temperature $\beta=1/k_{B}T$, with $k_{B}$ and $T$ being Boltzmann's
constant and the system temperature, respectively. There is a rigorous
theoretical framework to treat systems in such a way, however, we
will here limit the discussion to the time-discrete cases, as it most
directly relates to the MSM framework. Time-continuous models discussed
elsewhere e.g. \cite{Buchete_2008}, have analogous results \cite{Prinz_2011}.

The object of interest here is a \emph{propagator}, $\mathcal{P}_{\tau}$.
The propagator is an 'integral operator,' that acts on a probability
density function, $p_{t}(\cdot)$, over -- in our case -- conformational
space and returns the resulting probability density function on the
same space after a time, $\tau$. Formally,
\[
p_{t+\tau}(x)=[\mathcal{\mathcal{P}_{\tau}}p_{t}](x)=\int p(x\mid x',\tau)p_{t}(x')\,\mathrm{d}x'
\]
where, $p(x\mid x',\tau)$ is the transition probability density function
from $x'$ to $x$ after a time $\tau$. If $p_{t}(x)$ is equal to
the equilibrium distribution (Boltzmann distribution), then $p_{t+\tau}(x)=p_{t}(x)=\mu(x)$.
In general, if we apply the propagator to some initial distribution
$p_{0}(x)$ infinitely many times we arrive at the distribution $p_{\infty}(x)=\mu(x)$.
In other words, the propagator describes how an initial condition,
$p_{0}(x)$, relaxes to equilibrium. 

This observation reminds us of an Eigenvalue problem, where the Boltzmann
distribution is a solution (Eigenfunction), with the corresponding
Eigenvalue 1. Indeed, the propagator has infinitely many Eigenfunctions,
$\boldsymbol{\phi}_{i}$, whose Eigenvalues are bounded $1>|\lambda_{i}|$
for reversible dynamics. Ergodic dynamics further ensure only one
Eigenfunction has Eigenvalue 1: namely the Boltzmann distribution
$\boldsymbol{\phi}_{1}=\mu$. 

The Eigenvalues of $\mathcal{P}_{\tau}$ are the auto-correlations
of the Eigenfunctions $\phi_{i}$, which follow single-exponential
decays $c_{\phi_{i}}(\tau)=\lambda_{i}=\exp(-\kappa_{i}\tau)$, as
$\mathcal{P}_{\tau}$ is first-order Markovian. $\kappa_{i}\geq0$
are exchange rates, and $1/\kappa_{i}$ is often referred to as an
implied time-scale (ITS) \cite{Prinz_2011}. We immediately notice
that the implied-timescale for $\mu$ is equal to $\infty$, which
is consistent with our understanding that the Boltzmann distribution
is stationary: it does not change with time under fixed conditions.
Simultaneously, this observation suggests that all $\lambda_{i}<1$
approach 0 for large $\tau$, meaning that they -- together with
their corresponding Eigenfunctions -- encode information about the
dynamics of our molecular system. These Eigenfunctions (for $|\lambda_{i}|<1$)
describe what regions of conformational space exchange, on the timescale
$1/\kappa_{i}$. The negative and positive signs of an Eigenfunction,
$\boldsymbol{\phi}_{i}$, defines two regions of conformation space
that are exchanging on the timescale $1/\kappa_{i}$. 

So, the more Eigenfunction-Eigenvalue pairs we know the more we know
about the ensemble thermodynamics ($\mu$) and dynamics ($\boldsymbol{\phi}_{i>1}$,$\lambda_{i>1}$)
of our system -- but how do we deal with the infinite amount of these
pairs? --- This question is key to a central assumption made when
using MSMs: we are only interested in small number, $M$, of pairs
that correspond to those with the $M$ largest Eigenvalues. The larger
the Eigenvalue the slower time-scale -- consequently, we focus our
attention on slow dynamics. Immediately, this focus makes a lot of
sense, since long timescales often are associated with biological
function, including allosteric regulation and protein-protein binding.
Simultaneously, long-time scales remain challenging to study with
unbiased MD compared to fast dynamics. The success of this approach
lies in how representative the $M$ largest Eigenvalue-Eigenfunction
pairs are for the dynamics as a whole. Fortunately, for many systems
there are only a handful of Eigenvalues which are close to $1$, while
the rest are close to $0$.

Recall the dynamics in the continuous space is Markovian by construction.
To approximate the dynamics of a system from finite MD data, it is
an advantage to discretize conformational space. The Markov state
model (MSM) approach emerges naturally from this approximation. An
MSM aims to \emph{approximate} continuous space dynamics via a discrete
space jump-process on a partition of the configuration space into
$N$ disjoint segments. The discretization of the space and the $N\times N$
transition probability matrix, $T_{\tau}$, describing the 'jump-process'
constitute the approximation. Since $T_{\tau}$ is an approximation
of the continuous space dynamics, its Eigen\emph{vectors} and Eigenvalues
will --- if properly built -- approximate their corresponding quantities
in the continuous space dynamics. The Eigen-decomposition of $T_{\tau}$,
takes the form
\begin{eqnarray}
T_{\tau} & = & \sum_{i=1}^{N}\lambda_{i}\boldsymbol{l}_{i}^{\top}\mathbf{r}_{i}
\end{eqnarray}
where $\boldsymbol{l}_{i}$ and $\boldsymbol{r}_{i}$ are orthonormal
left and right Eigenvectors, respectively. The left Eigenvectors are
given by $\boldsymbol{l}_{i}=\boldsymbol{\mu}\circ\boldsymbol{r}_{i}$,
and $\circ$ is the element-wise product between two vectors. In this
expression, we see more explicitly how Eigenvectors with smaller numerical
Eigenvalues (faster time-scales) contribute less numerically to the
transition probability matrix.

\noindent\textbf{Key message: }The full-space dynamics contain essential thermodynamic
and kinetic information we need to characterize a protein-protein
encounter process. How well we reduce the full-space into a set of
discrete states controls the quality of our model. A sound reduction
of the full-space minimizes the error of the Eigenfunction corresponding
to the largest Eigenvalues of $\mathcal{P}_{\tau}$.%

\begin{figure}
\includegraphics[width=1\textwidth]{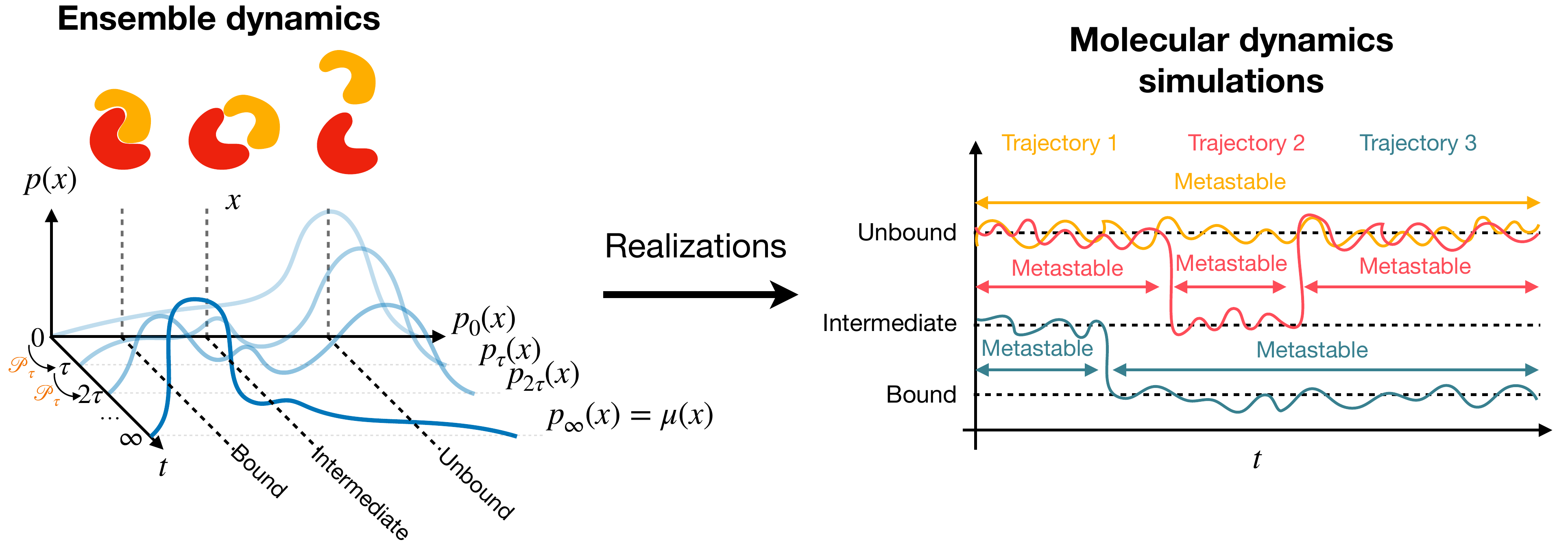}

\caption{Illustration of the relationship between the ensemble view of dynamics
and the individual realizations we obtain from molecular dynamics
simulations. \label{fig:Illustration-of-the}}

\end{figure}

\section{Strategies for MSM estimation, validation, and analysis}

As we saw above, building MSMs relies on discretizing conformational
space into N disjoint segments. These segments need to provide a good
basis for approximating the Propagator Eigenfunctions to ensure we
achieve the best possible approximation of the full space dynamics'.
The dimensions of the continuous space dynamics are many, for all
but the simplest system, so it is not practical to place a fine grid
on all dimensions. Placing such a grid would require enormous computer
memory and simulation data to be successful. We are facing what in
statistics is called the curse of dimensionality. In practice, building
a MSM involves a sequence of four steps \cite{Wehmeyer_2019},
\begin{itemize}
\item Featurization -- selecting a suitable representation of the molecular
system
\item Dimension reduction -- reducing the representation of the molecular
system
\item Clustering -- discretization of the representation
\item Transition matrix estimation -- estimation of the MSM,
\end{itemize}
each of these steps naturally introduces a substantial number of possible
modeling decisions. The following sections outline successful principles
to enable effective decision-making for all these steps.

\subsection*{Variational approach for conformational dynamics and Markov processes
(VAC and VAMP)}

When we build MSMs, we express the molecules' thermodynamic and kinetic
properties on a discrete set of disjoint states. Adopting this strategy
means that we approximate the eigenfunctions using a combination of
indicator functions -- functions that return one if we in a certain
area of configuration space and zero everywhere else. However, this
is just one way of approximating the Eigenfunction, and we are free
to approximate them with any function we like. The variational approach
for conformational dynamics (VAC) \cite{N_ske_2014,No__2013} gives
us a principle to select the function that best approximates a molecular
system's slow dynamics from a set of trial functions. Here, I briefly
outline the idea -- more detailed treatments are available elsewhere.

VAC uses that the Eigenvalues of $\mathcal{P}_{\tau}$ are bounded
and the Eigenfunctions form an orthonormal basis. Consequently, if
$f_{\alpha}$ is an approximation of the $\alpha$'th Eigenfunction
of $\mathcal{P}_{\tau}$ the auto-correlation is given by
\[
c_{\tau}(f_{\alpha})=\int f_{\alpha}(x)\mu^{-1}(x)\mathcal{P}_{\tau}f_{\alpha}(x)\,\mathrm{d}x\leq\lambda_{\alpha}
\]
where the equality holds if and only if $f_{\alpha}(x)$ is \emph{exactly}
the $\alpha$'th Eigenfunction of $\mathcal{P}_{\tau}$. Hence the
variational principle tells us that we will always approximate the
auto-correlation of an exact Eigenfunction $\phi_{\alpha}(x)$ from
below. Practically, this means we can devise algorithms to approximate
a set of orthonormal approximations of the Eigenfunctions of $\mathcal{P}_{\tau}$. 

A more general Variational approach for Markov processes (VAMP) \cite{Wu_2019},
extends VAC to non-reversible dynamics, non-equilibrium data, and
allows us to define scores which can be used for hyper-parameter optimization,
cross-validation, and model selection when building MSMs. These VAMP-scores
summarize the auto-correlations on a set of basis functions (features),
which best approximate the underlying dynamics, and therefore how
well they represent slow dynamics. We can use the VAMP-scores at every
step of the MSM building process to evaluate how well our modeling
decisions will allow us to represent the slow dynamics of a molecular
system.

\subsection*{Feature selection }

To facilitate the estimation of MSMs, we will need to arrive in a
sufficiently low-dimensional space to allow effective discretization.
However, the space has to include sufficient detail to capture the
interesting slow processes in our dataset. Fortunately, we frequently
have a clear idea of what kinds of processes we are interested in
resolving, or more specifically, what features we are not interested
in resolving. For example, in many cases, we are not directly interested
in studying the influence of solvation or the rotational and translational
motion of the solutes. This focus leaves us with studying different
internal coordinates or inter-molecular coordinates when selecting
features for building MSMs. These internal coordinates -- or \emph{features}
-- typically include contacts, distances, angles, and torsions between
atoms or atom groups.

While the considerations outlined above refines our choice of possible
structural features to use in our model building, it still leaves
open an enormous set of potential structural features. To further
narrow down this ambiguity, there are two different strategies: 
\begin{enumerate}
\item Manual feature selection by selecting features based on chemical,
biological, or physical insights which give us some information about
possible slow processes 
\item Algorithmic feature selection strategies. 
\end{enumerate}
It is difficult to approach the first strategy in a general and systematic
way. Typically, this strategy involves manually refining the selection
of features such that the model is robust and provides the necessary
predictive and descriptive power envisaged for the project. The second
approach is typically more systematic and generalizable and will normally
be the best choice if we know little about the system beforehand.
Several methods provide automated feature selection specifically designed
with MSM building in mind: Scherer, Husic \emph{et al.} illustrate
use of VAMP in this respect \cite{Scherer_2019}, and Chen \emph{et
al.} use a genetic algorithm based method for feature selection \cite{Chen2018}.
The former method works directly on the features, whether the latter
approach relies sub-sequent modeling steps to evaluate the selected
features. Therefore, the latter method is susceptible to confounding
factors limiting when evaluating the quality of a set of features.

\subsection*{Dimensionality Reduction }

Usually, pre-selecting several features (distances/contacts, angles,
features etc) using the strategies outlined above is insufficient
to sufficiently reduce the space to enable effective discretization
of the conformational space. Alternatively, we may not know much about
the system before starting our analysis, and we may want to identify
structural features that characterize the molecular dynamics well.
To face this problem, we can use dimensionality reduction techniques.
These methods remove dependencies in the input data through linear
(or non-linear) combinations learned utilizing a range of different
optimality criteria, thereby allowing us to represent the original
data in a lower-dimensional space while keeping the optimality criteria
used as small as possible. Dimensionality reduction techniques have
their origin in machine learning and statistics in a branch which
is now broadly referred to as unsupervised learning.

In the context of MSM, principal component analysis (PCA) \cite{Sittel_2014,Garc_a_1992,Ichiye_1991,de_Groot_2001}
and time-lagged independent component analysis (TICA) \cite{P_rez_Hern_ndez_2013,Schwantes_2013,Schwantes_2015}
are the most widely used, limiting the discussion here to these two.

PCA seeks to define a linear projection,
\[
\mathbf{Y}=\mathbf{X}\mathbf{W}
\]
of the set of input features, $\mathbf{X}\in\mathbb{R}^{N_{\mathrm{numframes}}\times N_{\mathrm{numfeats}}}$,
to \emph{maximize the variance} of each of the dimensions of $\mathbf{Y}\in\mathbb{R}^{N_{\mathrm{numframes}}\times N_{\mathrm{reduceddim}}}$,
by learning $\mathbf{W}\in\mathbb{R}^{N_{\mathrm{numfeats}}\times N_{\mathrm{reduceddim}}}$,
subject to an orthonormality constraint on the columns of $\mathbf{W}$
to ensure each dimension in $\mathbf{Y}$ is uncorrelated and normalized.
Consequently, PCA gives us a new set of features that best capture
our input features' variance and is an appropriate choice if we are
interested in studying processes characterized by large-scale structural
fluctuations.

TICA similarly seeks to find a linear projection as for PCA. However,
instead of maximizing the variance, TICA uses the variational principle
of conformational dynamics to determine projections with the slowest
auto-correlation. Consequently, TICA is the appropriate choice if
slow motions are of interest when studying a molecular system. Recall,
slow dynamics is what consititute the dominating part of the propagator.
Practically, we compute TICA by solving the generalized Eigenvalue
equation (subject to appropriate normalizations),

\[
\mathbb{C}_{\tau}\mathbf{w}_{i}=\mathbb{C}_{0}\lambda_{i,\tau}\mathbf{w}_{i}
\]
where $\mathbb{C}_{\tau}=\frac{1}{N_{\mathrm{numframes}}-\tau}\mathbf{X_{:\mathrm{numframes}-\tau}^{\top}}\mathbf{X}_{\tau:}$
and $\mathbb{C}_{0}=\frac{1}{N_{\mathrm{numframes}}}\mathbf{X^{\top}}\mathbf{X}$
are the time-lagged and instantenous covariance matrices, respectively.
$\mathbb{C}_{\tau}$ computes the covariance between features spaced
in time by $\tau$, and the indices $:\mathrm{numframes}-\tau$ and
$\tau:$ mean all but the last $\tau$ frames and the all but the
first $\tau$ frames, respectively. At this point $\tau$ is an integer
with a time-unit of the spacing interval between the frames in your
MD trajectory data. We can use the \emph{independent components},
$\mathbf{w}_{i}$, which solve this equation and correspond to the
largest eigenvalues $|\lambda_{i,\tau}|<1$ to project the data on
to a lower-dimensional space, $\mathbf{y}_{i}=\mathbf{X}\mathbf{w}_{i}$,
which conserve the slowest dynamic modes in the system. We use the
total kinetic variance $\varsigma_{\tau}^{2}$ to quantify how much
dynamic is preserved in the $d$-dimensional projection ($d<N_{\mathrm{numfeats}}$),

\[
\varsigma_{\tau}^{2}=\frac{\sum_{i=2}^{d}\lambda_{i,\tau}^{2}}{\sum_{j=2}^{N_{\mathrm{numfeats}}}\lambda_{j,\tau}^{2}}.
\]
Both PCA- and TICA-based dimension reduction methods are part of the
major MSM software packaged PyEMMA \cite{Scherer2015,Wehmeyer_2019}
and MSMBuilder \cite{Harrigan_2017}.

Recent surveys discuss the use of non-linear dimensionality reduction
techniques in the context of MSM estimation. While promising, these
methods have not seen broad adoption so far.

\subsection*{Clustering }

MSMs rely on discretizing the configurational space into disjoint
configurational states -- \emph{micro-states}. Clustering is the
step where the grouping of molecular configurations into discrete
states happens. The most commonly applied algorithm towards this purpose
is $k$-means clustering, yet several other methods perform this task
with a variety of different strategies \cite{Keller_2010}. As for
feature selection, we can use VAMP scores and cross-validation to
evaluate our clustering quality. Below, I expand on other considerations
which are important when clustering states when studying protein-protein
encounters.

\subsection*{Model Estimation and Validation}

Following clustering, we can assign every molecular configuration
to a Markov state. Trajectories now realize a jump-process on a set
of, $N$, discrete states, each of the states are connected back to
a molecular configuration. We call these \emph{discrete trajectories},
$D=\{d_{1},\dots,d_{M}\}$. The task of estimating a Markov state
model corresponds to computing the most likely transition probabilities,
$t_{ij,\tau}$ between any two states $i$ and $j$ after a lag-time
of $\tau$. Recall, we assume the dynamics are Markovian, so the \emph{likelihood}
of observing our data, $D$, is equal to the product of all the transition
probabilities,
\begin{eqnarray*}
\ell(T_{\tau}\mid D) & = & \prod_{d\in D}p(d[\tau]\mid d[0],\tau)p(d[2\tau]\mid d[\tau],\tau)\dots p(d[M]\mid d[M-\tau],\tau)\\
 & = & \prod_{d\in D}t_{d[0]d[\tau],\tau}t_{d[\tau]d[2\tau],\tau}\dots t_{d[M-\tau]d[M],\tau}\\
 & = & \prod_{ij}t_{ij,\tau}^{C(\tau)_{ij}}
\end{eqnarray*}
$C(\tau)$ is the count matrix where each element $C(\tau)_{ij}$
is the number of transitions between states $i$ and $j$, with a
time-lag, $\tau$, observed in all the trajectories, $D$. Estimating
a MSM then corresponds to finding the transition probabilities, given
the observed transition counts in the count matrix, $C(\tau)$. We
can either do maximum likelihood estimation\cite{Bowman_2009,Prinz_2011},
or Bayesian posterior sampling of the transition probabilties \cite{Trendelkamp_Schroer_2015,No__2008}.
The first approach gives us the one most likely model whereas the
latter approach gives us a distribution of models which we can use
to compute properties as well as their uncertainties. 

Major MSM software packages implement algorithms to perform inference
via either mode, with options to enforce constraints such as detailed
balance \cite{No__2008,Trendelkamp_Schroer_2015} or a fixed stationary
distribution \cite{Trendelkamp_Schroer_2013}. As outlined above,
the detailed balance constraint ensures a reversible MSM is estimated
and reduces the number of degrees of freedom to be estimated. Adding
constraints to the estimation when possible is often desirable, as
it may increases robustness of the results.

Choosing the lag-time when building a MSM decides the effective time-resolution
of the resulting model \cite{Husic_2017}. Consequently, we want to
keep this number small, too preserve as much of the information in
our data as possible. However, since we reduce a high-dimensional
space down to a lower-dimensional one to enable discretization, there
is no guarantee that the projected dynamics will be Markovian at short
lag-times \cite{Feng_2015,Su_rez_2016,Su_rez_2020}. We check the
'Markovianity' of the projected dynamics by computing the ITS as a
function of lag-time and ensuring no systematic change in the ITS
as a function of lag-time considering the statistical uncertainty.
A good choice of lag-time is then one which as short as possible,
yet shows now significant change in the ITS when increased or decreased
slightly. This analysis is typically facilitated by an ITS plot, showing
the ITS as a function of lag-time.

Having selected an appropriate lag-time, we can test the resulting
MSM for self-consistency with the simulation data via the Chapman-Kolmogorov
(CK) test \cite{Prinz_2011,No__2009}. This test makes use of the
time-discrete Chapman-Kolmogorov equation
\[
T_{k\tau}=T_{\tau}^{k}
\]
which predicts that the transition probabilities of a model estimated
with lag-time $k\tau$ should be equal to the transition probabilities
of a model estimated with lag-time $\tau$ to the power of $k$. We
typically visualize the CK test by comparing the values on either
side of the equation with error bars as a function of integer multiples
of the MSM lag-time. As for the ITS analysis, we here aim to see agreement
within statistical uncertainty. Usually, only a reduced set of states,
or a coarse-grained model, is used to facilitate analysis.

\subsection*{Spectral gaps and Coarse-graining}

It is not uncommon that MSMs end up having hundreds or thousands of
micro-states. The large number of micro-states helps us bring down
the error when approximating Eigenfunctions. However, it can stifle
the subsequent analysis. Consequently, we often coarse-grain the MSMs
into a handful of meta-stable macro-states, which summarize the slow
dynamics. Coarse-graining here should not be confused with coarse-grained
simulations, where beads represent multiple atoms. We have to decide
how many states, and that number may not be evident from the start.
In many cases, we can use the spectral gap in the Eigenvalue spectrum
of the MSM to decide on how many states we need to coarse-grain a
MSM to, to ensure we represent the slow dynamics. 

Suppose we sort the Eigenvalues-Eigenvectors pairs of a MSM by the
amplitude of the Eigenvalue, and plot them. In that case, we often
see one or more drops in the amplitude with increasing index (decreasing
Eigenvalue). These drops are spectral gaps and pin-point separations
between fast and slow dynamics in the molecular system represented
by the MSM. We can use these spectral gaps to decide on how many states
to use for a coarse-graining, as every Eigenfunction specifies what
two regions of conformation space are exchaning on the ITS which can
be computed from the Eigenvalue. Consequently, if we have $n$ Eigenvalues
which are less than 1 above a spectral gap, a $n+1$ state coarse-graining
will be appropriate. 

Perron Cluster-Cluster Analysis (PCCA) \cite{Deuflhard_2005,Kube_2007}
is a method that groups together micro-states based on the structure
of the Eigenvectors of a MSM is the most common way to identify important
meta-stable macro-states sampled during molecular dynamics simulations.
Two related algorithms, PCCA+ and PCCA++, find an optimal linear transformation
of the Eigenvector coordinates onto a probability simplex \cite{R_blitz_2013}.

Hidden Markov state models (HMM) are an alternative to both MSMs and
PCCA \cite{No__2013b}. HMMs avoid the assumption of MSMs of Markovian
dynamics in the reduced space by estimating a \textquotedblleft hidden\textquotedblright{}
Markov chain observed indirectly via the trajectory data on the discrete
micro-states. A HMM therefore estimates a transition probability matrix,
and an ``emission matrix,'' $E$, the first matrix is responsible
for modeling the dynamics, and the latter models the observation process:
given we are in hidden state $i$ we will be in microstate $j$ with
probability $p(j\mid i)=E_{ij}$. Consequently, the emission matrix
tells us what states exchange rapidly, given we are in a specific
meta-stable configuration. We can use this to simplify the many states
into just a few states. HMMs have a range of other theoretical advantages
but are also more challenging to estimate than MSMs. There are other
alternatives to defining lower-dimensional models to facilitate analysis
of slow dynamics in terms of a few meta-stable states. However, their
performance in the context of protein-protein encounters is currently
unknown \cite{Hummer2014,Gerber2018}. 

\subsection*{Adaptive and enhanced sampling strategies}

The quality of the molecular dynamics simulation data ultimately determines
the quality of the estimated MSMs. Here, quality means the number
of transitions sampled between configurational states of interest
for the molecular system. An advantage of MSM analysis is that we
do not necessarily need to sample transitions between all states of
interest in every trajectory but sample only a subset of the possible
transitions. However, in practical cases, we still have to make the
most of limited resources -- blindly or naively running numerous
simulations may not be the most effective.

Adaptive sampling strategies (semi) automatically decide how multiple
simulations run in parallel and over several \textquotedblleft epochs.\textquotedblright{}
These strategies have to balance exploration and exploitation: sampling
new states and refining sampling statistics between previously visited
states \cite{Zimmerman_2015,Hruska_2018,Bowman_2010}. Several groups
have proposed strategies using different assumptions about what is
important to characterize molecular systems \cite{Zimmerman_2015,Plattner_2017,Doerr_2016,Zimmerman_2018,Wang_2017,Hruska_2020,Doerr_2014,2002.12582}.
A complementary set of strategies aim to sample transitions between
known states \cite{Chong_2017,Zuckerman_2017,Ahn_2020}. However,
due to the relatively high computational cost of studying protein-protein
encounters, these methods are yet to be compared in rigorous benchmarks.

Enhanced sampling methods bias molecular dynamics simulations intending
to speed up sampling processes of interest, such as protein-protein
binding and unbinding \cite{Zhou,Laio_2002,Grubm_ller_1995,Huber_1994,Kumar_1992}.
Unfortunately, introducing the right biases to enhance the sampling
of a process of interest remains a labor-intensive process. Nevertheless,
methods are available to recover stationary properties from biased
simulations, yet proving more difficult for dynamic properties. However,
combining unbiased and biased simulation data via recent MSM estimation
techniques can significantly improve the estimated models' robustness
\cite{Chodera_2011,Wu_2016,Mey_2014,Rosta_2014,Stelzl_2017,Paul_2017}.
In section \ref{sec:Successful-examples-from} we highlight the successful
use of adaptive and enhanced sampling techniques to study protein-protein
binding-unbinding modeling.

\subsection*{Practical consideration for studying protein-protein encounters}

The procedures outlined above generally apply to molecular systems.
However, there are additional aspects that are critical to be mindful
of when modeling protein-protein encounters. 

We can use macroscopic variables, including concentration, temperature,
pressure, and mutations, to control the molecular system's ensemble,
including the population of bound and unbound states and their kinetics
of exchange. As we have discussed, following experimental observables
as a function, these variables allow us to quantify essential properties
such as affinities, rate constants, and structural information about
the complex formation process.

Computationally, we often have to settle on a single -- or a few
-- macroscopic setting(s) of variables to study. This limitation
is due to the large computational requirements associated with sampling
each condition, even using the advanced simulation strategies, including
those outlined above. A notable exception is the temperature, which
is leveraged in enhanced sampling techniques to improve sampling efficiency.
When analyzed together with regular MD simulation data using appropriate
statistical estimators, they may improve MSM estimation. However,
using these data on their own makes it challenging to get insights
about exchange kinetics between conformational states.

The primary differences we face when studying protein-protein encounters,
compared with studying the molecular dynamics in a single protein
molecule, are stoichiometry and concentration. Practically, the simulation
volume is limiting: when we increase the volume, we need to simulate
larger systems, usually comprised of an increasing number of water
molecules. This fact makes simulations with high protein concentrations
the only computational viable strategy currently.

A high concentration in molecular dynamics has some practical consequences
which may make it practically difficult to study certain mechanisms
of protein-protein encounters. Let us consider a case of the conformational
selection mechanism, where a low-population state $A^{*}$ of unbound
protein $A$ binds the protein $B$ to form the complex $A^{*}:B$
in the following reversible chemical kinetic relation

\[
A\rightleftarrows A^{*}+B\rightleftarrows A^{*}:B
\]
where $k_{A\rightarrow A^{*}}\ll k_{A^{*}\rightarrow A}$ with both
rates being independent of concentration. We assume the protein $B$
does not undergo conformational changes which perturb this relation
directly. The on-rate, $k_{A^{*}+B\rightarrow A^{*}:B}$, is proportional
to the protein concentration and the population of the unbound state
of $A^{*}$, so as concentrations increase, the probability of observing
binding events increase. In an alternative binding mechanism (induced
fit) binding happens before conformational change in the protein $A$,

\[
A+B\rightleftarrows A^{*}:B.
\]
Here, the on-rate (the rate of binding), $k_{A+B\rightarrow A^{*}:B}$,
depends on the protein concentration and the population of the highly
populated state of protein $A$. In many reported cases both mechanisms
are possible, consequently, we seek to understand the balance of these
two mechanisms -- more generally, we seek to characterize the association-dissociation
path ensemble \cite{E__2006,Metzner_2009}. However, since we are
at high concentrations we may have $k_{A^{*}+B\rightarrow A^{*}:B}\ll k_{A+B\rightarrow A^{*}:B}$,
and we may even have $k_{A\rightarrow A^{*}}\ll k_{A^{*}\rightarrow A}\ll k_{A+B\rightarrow A^{*}:B}$.
So with only finite MD simulation data, we may severely under-sample
or completely miss certain mechanisms, even if they are important.
In other words, high protein concentrations in MD simulations may
increase the free energy of the unbound state to the point where the
association is barrier free, and the unbound state is not meta-stable
\cite{Weikl_2014}.

More concretely, given the competition between these mechanisms, sampling
the induced-fit mechanism is much more likely than sampling the conformational
selection mechanism. Even conformational sampling of protein $A$
is much less likely than sampling binding via induced fit. Practically,
these conditions mean that we will have an intrinsic preference to
observe a certain biophysical binding mechanism and may over-sample
mechanisms that are not relevant at physiological protein concentrations,
including unspecific binding events. As a result, we would need to
acquire more simulation data to ensure statistically sufficient sampling
of alternative binding mechanisms and conformational mixing of the
unbound states.

The fast on-rates at high concentrations may also influence our ability
to distinguish unbound and bound states automatically. The time-scale,
$t_{i}$, of a process, $i$, between states $a_{i}$ and $b_{i}$
is depends on the geometric average of the rates of the forward and
backward process $t_{i}=\frac{1}{\kappa_{i}}=\frac{1}{k_{a_{i}\rightarrow b_{i}}+k_{a_{i}\rightarrow b_{i}}}$,
which is numerically dominated by the faster (larger) rate. As a result,
the binding-unbinding process's time-scale will be fast. Therefore,
dimension reduction techniques may not resolve it as an important
process. Consequently, a MSM based on a clustering defined only in
this space will miss the process altogether. However, we can overcome
this problem by explicitly separating bound and unbound states, such
as a molecular feature that clearly distinguishes the unbound and
bound states.

The chosen forcefield model may significantly affect the sampled binding
mechanisms and may be prone to deficiencies such as strong unspecific
binding. Although efforts continuously improve these forcefield models
address their outstanding limitations, we often do not know how well
it will represent a new system of interest before we start simulations.
In the next section, we discuss strategies to validate MSMs built
using potentially imperfect forcefield models and possibly overcome
some of the limitations.

\subsection*{Analysis of the association-dissociation path ensemble}

The 'mechanism' of binding is ultimately governed by the statistical
distribution of different paths from the unbound state to the bound
state. The importance of the different paths between the unbound and
bound states is governed by the flux along that path. Transition path
theory (TPT) \cite{E__2006,Metzner_2009,Prinz_2011b,No__2009} provides
us with a theoretical framework through which we can compute reactive
flux-matrices from MSMs of protein-protein encounters. The 'reaction'
here refers to the transition from a set of 'reactant states' (unbound)
$A$ to a set of 'product states' $B$ (bound). TPT gives us tools
to assign all intermediate states $I$ (not bound, and not unbound),
committor probabilities, $q^{+}$ and $q^{-}$, which tells gives
us the probability of reaching the $B$ before $A$ from an intermediate
state $i\in I$ via forward committor $q_{i}^{+}$ and vice-versa
for the backward committor $q_{i}^{-}$. We can further use this framework
to compute mean first passage times (MFPT), for example the average
on- and off-rates, as well as dissect all the possible pathways from
unbound to bound states. TPT is therefore an important tool for analysis
of MSMs, in particularly when we want to understanding a specific
process. Major MSM softwares implement TPT analyses and plotting functions
to visualize the results \cite{Wehmeyer_2019,Scherer2015,Harrigan_2017}.

\begin{figure}
\includegraphics[width=1\textwidth]{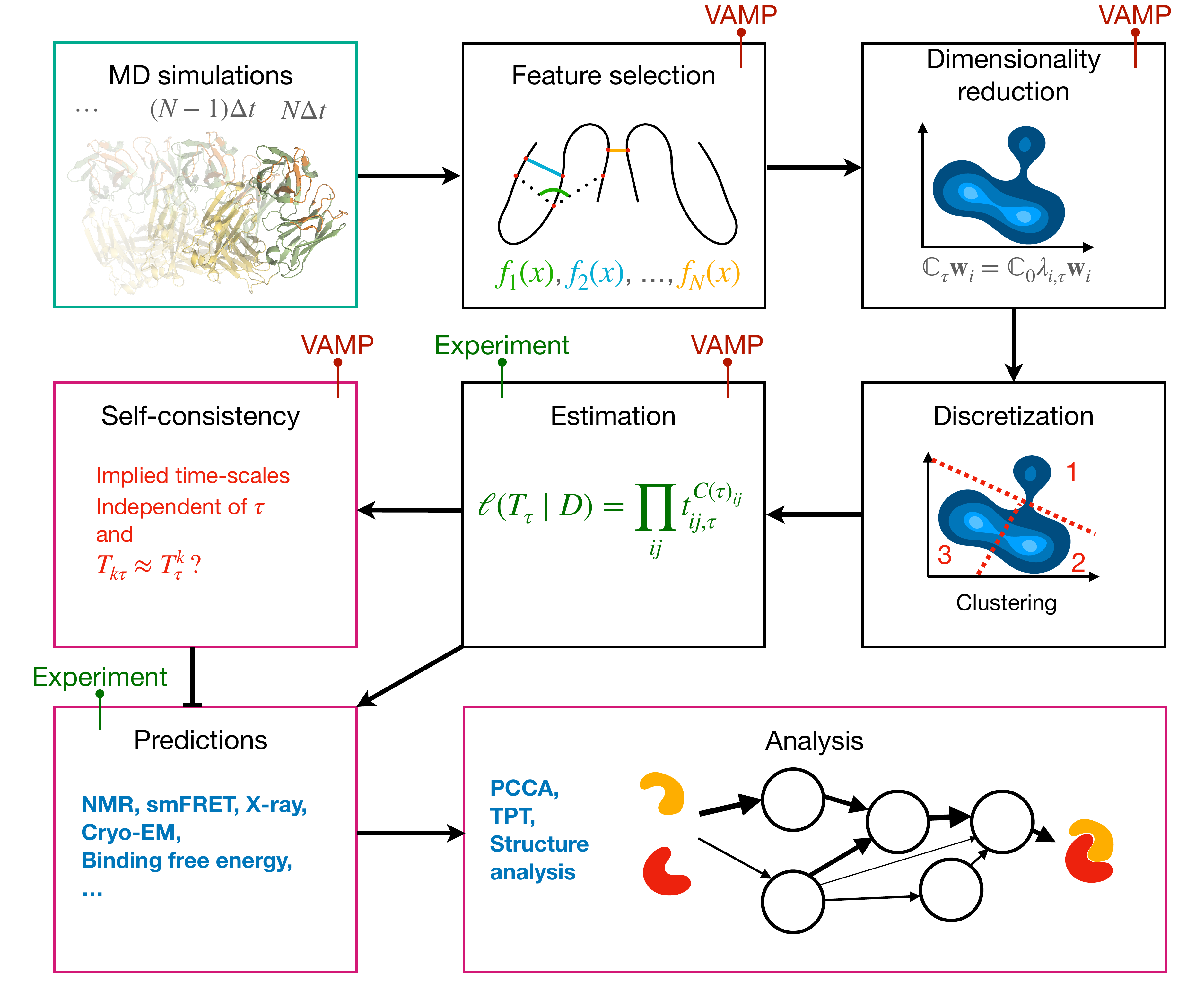}

\caption{Flow-chart of Markov state modeling from molecular dynamics simulations
to final model and analysis. Boxes are colored to indicate data collection
(cyan), data processing and model estimation (black), and analysis
and validation (magenta). I highlight steps which may benefit from
specific techniques or experimental data with colored pins.}

\end{figure}

\section{The connection to experiments }

In favorable cases, appropriately validated MSMs predict molecular
mechanisms with high temporal and spatial resolution. These insights
can, of course, guide our understanding of important molecular phenomena
associated with, for instance, protein-protein binding. However, so
far, we have only discussed validation as statistical self-consistency
and minimizing projection errors (optimizing variational scores).
Since we generate the simulation data that we use to drive the estimation
MSMs with imperfect classical empirical force field models, agreement
with experimental data is not a given. In this section, I will outline
how we can predict important biophysical observables to check for
agreements, the limitations of these comparisons, and how we may integrate
experimental data into the estimation of MSMs using the Augmented
Markov model framework, to bring experiment and simulation into alignment.

\subsection{Experimental observability, forward models, and errors.}

What is an observable? -- In our context, an experimental observable
is a function of state; that means a function of the configurations
adopted by a molecular system at specific experimental conditions.
The definition encompasses both bulk experiments, where a very large
number ($\sim10^{23})$ of copies of identical systems cumulative
signal is measured, and experiments where time-resolved trajectory
signals from single molecules are measured. The manifestation of a
particular observable is described by a physical model, $f(\cdot)$,
describing the relationship between a configuration of the system,
$x$, and the observed signal, $o$. Simple examples of $f$ may be
a ruler measuring the Euclidean distance between two atoms in a molecule,
or a function computing the potential energy of the system configuration,
$x$. 

In a \emph{stationary }bulk experiment at equilibrium we measure an
expectation value of $f(\cdot)$ under the Boltzmann distribution,
\begin{eqnarray*}
\langle O_{f}\rangle & = & \mathcal{Z}^{-1}\int f(x)\exp(-\beta U(x))\,\mathrm{d}x\\
 & = & \int f(x)\mu(x)\,\mathrm{d}x\\
 & = & \mathbb{E}_{\mu(x)}[f(x)].
\end{eqnarray*}
We use $p(x)$ as short-hand for the normalized Boltzmann distribution
for a given $\beta$, and the brakets to denote ensemble averages.

In an ergodic, \emph{dynamic }bulk experiments at equilibrium we measure
the auto-correlation,
\begin{eqnarray*}
\langle O_{f}(0)O_{f}(\Delta t)\rangle & = & \int\int f(x')p(x'\mid x;\Delta t)p(x)f(x)\,\mathrm{d}x\mathrm{d}x'\\
 & = & \mathbb{E}_{p(x)}[\mathbb{E}_{p(x'\mid x,\tau)}[f(x')f(x)]]
\end{eqnarray*}
where $p(x'\mid x;\Delta t)$ is the transition probability. Note
that we may analogously define cross-correlation experiments by using
two different models for the observables,

\begin{eqnarray*}
\langle O_{f_{a}}(0)O_{f_{b}}(\Delta t)\rangle & = & \mathbb{E}_{p(x)}[\mathbb{E}_{p(x'\mid x,\tau)}[f_{b}(x')f_{a}(x)]].
\end{eqnarray*}
Some experimental setups will allow us to initialize an ensemble in
a non-equilibrium ensemble $p_{0}$ and follow the relaxation process
back to equilibrium \cite{Prinz_2011b,No__2009}. Such experiments
include pressure- and temperature-jump, as well as, stopped-flow.
Cross-correlation functions measured in such relaxation experiments
can be expressed as,
\begin{eqnarray*}
\langle O_{f_{a}}(0)O_{f_{b}}(\Delta t)\rangle_{p_{0}} & = & \int\int f_{b}(x')p(x'\mid x;\Delta t)\frac{p_{0}(x)}{p(x)}f_{a}(x)\,\mathrm{d}x'\mathrm{d}x.
\end{eqnarray*}
In single molecule experiments observables are followed over time
as trajectories, with some time resolution $\Delta t$, in its simplest
form:
\[
O_{f}=\{f(x(0)),f(x(\Delta t)),\dots,f(x(N\Delta t))\}.
\]

\paragraph{Sources of errors and uncertainty}

In a typical setting, we have some set of experimental data, which
may include any combination of the classes above, and we wish to compare
these observables to corresponding predictions made by our computational
model stationary and dynamic properties of the molecular system. Several
sources of uncertainty and error may arise in this setting that we
will have to be mindful of:
\begin{enumerate}
\item \textbf{Experimental noise (thermal noise, shot-noise, etc.)}\\
This category includes every stochastic noise contribution due to
limitations in the experimental setup, typically due to imperfections
in instrumentation measurements or sample (labeling) stability. In
many cases, theoretical analyses of experiments are available which
may help decide how this error should be modeled. 
\item \textbf{Systematic experimental errors/biases}\\
These errors and biases arise due to imperfect referencing --- in
the case of for example relative experimental measurements --- or
unknown, or imprecise, experimental conditions (temperature, concentrations,
pressure, etc) or fluctuations of these parameters during data acquisition.
This source of error is typically more challenging to systematically
model or perfectly compensate for and often require substantial knowledge
of the experimental setup.
\item \textbf{Systematic error in the computational model of the molecular
system dynamics}\\
Computational models of $p(x)$ and $p(x'\mid x,\Delta t)$ are typically
estimated using finite simulation data using from empirical forcefield
models. The quantitative agreement of these with experiment is ever
improving, however, still suffer significant systematic errors. The
errors may arise from the classical approximations made of quantum
mechanical interactions, to make simulations computationally tractable,
or other approximations. Our efforts to compare simulation models
to experiments are typically driven by the recognition of these issues,
and a desire to understand the limitations and merits of a given moded.
Another systematic error encompassed in this section is the sampling
error, where we simply do not have enough simulation data to accurately
estimate $p(x)$ and $p(x'\mid x,\Delta t)$. 
\item \textbf{Modeling error of observable functions $f(\cdot)$}\\
Like simulation models, the forward prediction of instantaneous (time-independent)
experimental observables approximate complicated experimental setups
or quantum mechanical phenomena in a computationally efficient manner.
In many cases, quantifying errors and biases in these models is challenging.
\end{enumerate}
\pagebreak{}

\subsection{Predicting experimental observables using MSMs}

The expressions given above for the experimental observables are general
for any ergodic, Markovian dynamics, at, or relaxing to, a time-invariant
equilibrium state. However, for molecular systems, these expressions
involve intractable integrals over the configuration space. Fortunately,
for MSMs, these integrals simplify to standard linear algebra operations,
which we can compute efficiently.

The discretization of configurational space, $\mathcal{S}=\{S_{1},\dots,S_{N}\}$,
associated with the MSM, leads to a discretization of the instantaneous
experimental observable (predicted via $f(\cdot)$) as the vector
$\mathbf{a}\in\mathbb{R}^{N}$, with elements,
\[
a_{i}=\text{\ensuremath{\frac{1}{\int_{x\in S_{i}}p(x)\,\mathrm{d}x}}}\int_{x\in S_{i}}p(x)f_{a}(x)\,\mathrm{d}x\approx\frac{1}{N_{S_{i}}}\sum_{j\in S_{i}}f(x_{j})
\]
where $S_{i}$ is configurational space segment, or its finite sample
approximation with $N_{S_{i}}$ samples. We can extend this expression
to vector-valued observables. Since this discretization replaces a
function which takes on arbitrary real valued numbers for different
conformations, by a piece-wise constant function, we hope to minimize
the variance of $f(x)$ within each $S_{i}$ to ensure a good approximation. 

We use the stationary distribution $\pi$, of the transition matrix
$T_{\tau}$, along with discretized feature vector $\mathbf{a}$ to
compute \emph{stationary bulk experimental observables} as,
\[
\langle O_{f_{a}}\rangle=\pi\cdot\mathbf{a}=\sum_{i=1}^{N}\pi_{i}a_{i}.
\]
The expression for \emph{dynamic bulk experiments}, can be expressed
using the transition matrix,

\begin{eqnarray*}
\langle O_{f}(0)O_{f}(N\tau)\rangle & = & \mathbf{a}^{\top}\boldsymbol{\Pi}T(\tau)^{N}\mathbf{a}
\end{eqnarray*}
where $\boldsymbol{\Pi}=\text{diag}(\pi)$ is a matrix with stationary
probabilities on the diagonal, and zeros elsewhere. $N$ is an integer
expressing time in multiples of the MSM lag-time $\tau$. Cross-correlation
experiments can similarly be expressed a
\[
\langle O_{f_{a}}(0)O_{f_{b}}(N\tau)\rangle=\mathbf{b}^{\top}\boldsymbol{\Pi}T(\tau)^{N}\mathbf{a}
\]
where $\mathbf{b}$ is defined as $\mathbf{a}$, but for a observable
predicted by $f_{b}(\cdot)$. In general, MSMs predict auto- and cross-correlation
function as mixture of exponential decays. We see this more directly
by considering the spectral decomposition of the transition matrix,
\begin{eqnarray*}
\langle O_{f}(0)O_{f}(N\tau)\rangle & = & \mathbf{a}^{\top}\boldsymbol{\Pi}\left(\sum_{i=1}^{N}\lambda_{i}^{N}\boldsymbol{l}_{i}^{\top}\mathbf{r}_{i}\right)\mathbf{a}\\
 & = & \sum_{i=1}^{N}\lambda_{i}^{N}(\mathbf{a}^{\top}\cdot\boldsymbol{l}_{i}^{\top})(\boldsymbol{l}_{i}\cdot\mathbf{a})\\
 & = & \sum_{i=1}^{N}\exp(-\frac{N\tau}{t_{i}})(\boldsymbol{l}_{i}\cdot\mathbf{a})^{2}\\
 & = & (\pi\cdot\mathbf{a})^{2}+\sum_{i=2}^{N}(\boldsymbol{l}_{i}\cdot\mathbf{a})^{2}\exp(-\frac{N\tau}{t_{i}}).
\end{eqnarray*}
Similar expression can be written down for cross-correlation and relaxation
experiments (Keller Prinz fingerprints). This simple form of the auto-
and cross-correlation functions from MSMs facilitates analytical expression
of several experimental observables including from NMR spectroscopy,
dynamic neutron scattering, and FRET spectroscopy.

\subsection{Integrating experimental and simulation data into Augmented Markov
models}

As mentioned above, systematic errors in the empirical force field
models used for molecular simulations to build MSMs lead to statistically
robust yet systematic errors in our predictions. Using the equations
above, we can quantify, but not remedy, these biases. A wealth of
methods have been introduced to bias MD simulations \cite{Olsson_2013,Olsson_2014,Bonomi_2016,Cavalli_2013,Pitera_2012,White_2014,Hummer_2015,Olsson_2015},
or reweight simulation data \emph{a posteriori} \cite{Olsson_2016b,Hummer_2015,Bottaro_2020,Olsson_2017,Brotzakis_2020,Ge_2018,Salvi_2016,K_mmerer_2020},
to match experimental data, using different inference philosophies
-- several excellent reviews discuss these approaches in more detail
\cite{Orioli_2020,Boomsma_2014,Bottaro_2018}. In the context of MSMs,
we already have simulation data available or are in the process of
adaptively acquiring it. Consequently, adopting an approach that would
alter our MD simulations' ensemble simulations is undesirable --
excluding the use of experimental data to bias simulations. On the
other hand, reweighing MD trajectories generally sacrifice the dynamic
information from our simulation data.

The augmented Markov models (AMM) \cite{Olsson_2017} framework allows
us to balance experimental and simulation data when building Markov
models of molecular kinetics. AMMs therefore achieve better agreement
with experimental data while preserving the dynamic information from
molecular simulations. To estimate AMMs the log-likelihood function
of MSMs augmented with a term to balance systematic discrepancies
between experimental and simulation data via a set of Lagrange multipliers
$\boldsymbol{\lambda}$, 
\begin{equation}
\ell(T_{\tau},\boldsymbol{\lambda}\mid\mathbf{C}(\tau),\mathbf{O},\boldsymbol{\sigma})\propto\sum_{ij}c_{ij}\log t_{ij,\tau}-\sum_{k}\frac{(\hat{m}_{k}-o_{k})^{2}}{2\sigma_{k}^{2}}\label{eq:AMM_loglik}
\end{equation}
where $t_{ij,\tau}$ is the $i,j$'th element of $T_{\tau}$, $c_{ij}$
the corresponding element in the count matrix $\mathbf{C}(\tau)$,
and $o_{k}$ and $\sigma_{k}$ are the $k$'th experimental observable
and its experimental uncertainty respectively. The prediction of the
experimental expectation value from $T_{\tau}$ is $\hat{m}_{k}=\mathbf{a}_{k}\cdot\hat{\pi}$
with
\[
\hat{\pi}_{i}=\frac{\pi_{i}\exp(\sum_{v}\lambda_{v}a_{v_{i}})}{\sum_{j}\pi_{j}\exp(\sum_{v}\lambda_{v}a_{v_{j}})}
\]
models the experimental Boltzmann distribution via a Maximum Entropy
perturbation of the simulation ensemble $\pi$ computed from $T_{\tau}$.
$\lambda_{v}$ is the $v$'th Lagrange multiplier corresponding to
experimental observable $o_{v}$ and its back-prediction for a Markov
state $i$ as $a_{v_{i}}$. Optimizing (eq. \ref{eq:AMM_loglik})
subject to detailed balance constraints yields an AMM.

We motivate the use of a Maximum Entropy perturbation as it provides
a model as close as possible -- in the Kullback-Liebler sense --
to the simulation ensemble. The critical assumption is therefore that
the simulation ensemble provides a reasonable starting point to model
the experimental data, including covering all meta-stable configurations
necessary accurately predict the experimental observables.

Other approaches similarly allow for the integration of experimental
data into MSM estimation. One approach enables the gradual adjustment
of MSM stationary distributions against target observables \cite{Rudzinski_2016}.
Matsunaga and Sugita present a method to integrate single-molecule
FRET data and molecular simulation using a step-wise HMM estimation
procedure \cite{Matsunaga_2018}. Brotzakis et al. propose a maximum
entropy and maximum caliber approach to reweigh trajectory ensembles
against bulk observables \cite{Brotzakis_2020}. 

Although the integration of experimental data and simulation data
has been an active area of research for several decades, several problems
remain open. In particular, some data still cannot be included in
the AMMs framework, including single-molecule FRET trajectories or
dynamic bulk observables.

\pagebreak{}

\section{Protein-protein and protein-peptide encounters}

\label{sec:Successful-examples-from}Several groups have reported
kinetic models of protein-protein and protein-peptide encounters using
molecular dynamic simulations and MSMs. As yet, tightly binding complexes
(small dissociation constants, $K_{D}$) with slow association-dissociation
kinetics dominate the literature, as they constitute the biggest challenge
for molecular simulations. While slow macroscopic kinetics, and large
free energy differences, characterize these systems, microscopically,
these protein-protein and protein-peptide encounters may happen via
multi-step processes. Consequently, we can sample rare events on the
seconds to minutes time-scale by connecting the bound and unbound
states sampling the much more likely transitions between intermediate
states. MSMs excel in cases such as these: transitions sampled between
intermediate steps along binding-unbinding paths can be combined into
a model that predicts the slow macroscopic dynamics of the full binding-unbinding
process inaccessible for direct simulation.

The first reported study of a full reversible protein-protein binding
study by all-atom molecular dynamic simulations was for the inhibitory
complex of ribonuclease barnase and barstar. The barnase:barstar complex
is an excellent benchmark system due to its extensive experimental
charaterization by multiple biophysical methods. Plattner \emph{et
al. }collected molecular simulations with 2 milliseconds of aggregate
simulation length \cite{Plattner_2017}. The data was distributed
between 1.7 milliseconds of independent simulations initiated from
dissociated states and 0.3 milliseconds using adaptive sampling. The
adaptive scheme allowed the authors to sample barnase:barstar association
with a few microseconds of aggregate simulation time, while the equilibrium
binding rate is tens of microseconds. Unbinding similarly benefitted
from adaptive sampling, with unbinding events being sampled in a few
hundred microseconds, while the equilibrium off-rate is expected to
be on the hours time-scale.The authors estimated a HMM to compute
the thermodynamics, kinetics, and important structural states of the
protein-protein encounter process. They compare predictions of macroscopic
thermodynamics and kinetic observables against experiment: binding
free-energy $12-19\,\mathrm{kcal}\cdot\mathrm{mol}^{-1}$ against
the experimental $16.8\,\mathrm{kcal}\cdot\mathrm{mol}^{-1}$ and
the dissociation rate $3\cdot10^{-6}-10^{-1}$ compared to the experimental
range of $8\cdot10^{-5}-5.0\cdot10^{\text{\textendash}4}\,\text{s}^{\text{\textendash}1}$.
The large uncertainties, illustrate how MSMs and HMMs quality, accuracy,
and precision for these quality critically rely on the number of binding
and unbinding events sampled in the aggregate simulation data. Nevertheless,
on-rate could be predicted with high accuracy and the most stable
state coincided with the crystallographic structure (pdb: 1BRS). Further,
perturbation theory allowed for accurate prediction of binding free-energy
changes upon mutation within stastistical uncertainty. The resulting
barnase:barstar HMM predicts a binding mechanism, where barstar can
associate to all points of barnase's surface, early intermediates
preferably binds opposite to the native binding groove, and late intermediates
states, and a 'trap' state, bind close to the binding grove, but in
non-native orientations. Later still in the process, complex passes
through late intermediates into a pre-bound, loosely bound, and then
finally the native bound state. The rate-limiting step is the pre-bound
bound state which is stabilized by electrostatic and hydrophobic interactions
between the two protein domains. 

Two studies report MSMs of protein-peptide encounters involving the
p53-antagonist MDM2 \cite{Paul_2017,Zhou_2017}. The first study investigates
one antagonistic pathway of MDM2 via its binding of the p53 transactivation
domain (TAD) \cite{Zhou_2017}. The study models this interaction
via a TAD peptide and uses extensive, unbiased molecular dynamics
simulations, as for the barnase:barstar study discussed above. The
second study instead reports the binding of MDM2 to an inhibitory
peptide PMI via integrating unbiased simulations and data from enhanced
sampling \cite{Paul_2017}. The first study, reports MSM that predicts
quantities with an accuracy comparable or worse than that observed
for the barnase:barstar case above. Qualitatively accurate on-rates,
yet off-rates do not agree with experiments -- this is likely due
to the relatively small data-set used here of 831 microseconds in
aggregate length and force-field errors. Nevertheless, the authors
can identify important structural states and investigate possible
binding mechanisms. Their model favors an induced-fit binding mechanism,
where TAD first binds MDM2 and then folds into the native complex
structure.

For the second study \cite{Paul_2017}, and in a follow-up study,
the authors use multi-ensemble Markov models (MEMMs) to quantitatively
predict binding thermodynamics and kinetics of MDM2 to the PMI peptide
with high precision. MEMMs define MSMs over multiple thermodynamic
states, such as those used in enhanced sampling techniques, including
replica-exchange and umbrella sampling. This approach's advantage
relies on fewer simulation data (approximately 102 microseconds of
Hamiltonian replica exchange and 500 microseconds of unbiased MD).
This disadvantage depends on designing an effective enhanced sampling
strategy for the system of interest, which may be challenging to achieve
without substantial trial-and-error and extensive human intervention.

To summarize, MSMs and related kinetic modeling approaches are currently
the only available strategy to gain microscopic insights into the
thermodynamics and kinetics of protein-protein and protein-peptide
encounters. These analyses can help us distinguish between different
binding mechanisms, and combined with perturbation approaches, qualitative
insights into the influence of point mutations impact the binding.
However, collecting sufficient data remains a serious challenge when
applying these methods in practice. At simulation rates of around
400-500 nanoseconds per day and GPU collecting millisecond sized datasets
may take GPU-years to complete. New adaptive sampling strategies and
the integration of experimental data and enhanced sampling simulations
may help lower demands on unbiased simulations. Nevertheless, studies
have focused on relatively small protein-protein and protein-ligand
systems and interactions with high affinity. Further improvement in
computing power, simulation, and analysis methods is needed to ensure
these analyses can benefit structural biology more broadly. Finally,
how these approaches will fare on low-affinity complex systems with
a less clear separation of time-scales also remains to be understood.

\section{Emerging technologies}

As we saw above, Markov state models are emerging as an important
tool in characterizing the thermodynamics and kinetics of protein-protein
encounters in ideal cases, providing detailed mechanistic models at
atomic resolution. We are steadily progressing towards better methods
for featurization, dimension reduction, clustering, and adaptive sampling
strategies; these advances contribute to minimizing the computational
and labor effort needed to build high-quality MSMs. However, we remain
reliant on access to state-of-the-art computing resources and, in
many cases, the extensive manual intervention of highly-skilled researchers.
We are further simulating increasing size scales poorly in terms of
simulation efficiency and the amount of simulation data needed to
build statistically sufficient models. A broad range of machine learning
methods is currently emerging that directly address the challenges
faced by MSMs.

Recall, the fundamental use of MSMs is to build a low-dimensional
approximation of an infinite-dimensional molecular dynamics operator.
This task relies on several pre-processing steps in sequential succession:
featurization, dimension reduction, clustering, and model estimation.
The success at each stage depends on the careful adjustment of hyper-parameters
against an optimality criterion. An error or sub-optimal choice made
early on in the sequence may negatively impact our final model's quality.
Yet, identifying and resolving such problems is often not straightforward
and relies on extensive testing and manual intervention. With VAMPnets,
Mardt et al. illustrate how we may, in principle, replace the entire
sequence of pre-processing steps and the model estimation by an artificial
neural network \cite{Mardt_2018}. The key idea is to input all-atom
coordinates into a neural network that outputs a categorical distribution
representing a conformation membership to $N$ metastable states.
The neural network's optimization objective function is a VAMP score
computed between pairs of simulation frames with a time-lag of $\tau$.
The neural network learns the complicated function from molecular
dynamics trajectories to a molecular kinetics model in a single step
through this procedure. Extensions of VAMPnets are already emerging
to improve data efficiency and impose further constraints on the modeled
dynamics.

The number of states a molecular system may potentially adopt grows
exponentially with its size. So, in addition to declining simulation
rates with system size, we must sample a much larger conformational
space. In other words, our simulations get slower, and we have to
simulate more to ensure statistically sufficient models. The latter
of these two problems arises from how we represent the meta-stable
states as global configurations. Consequently, we need to explicitly
account for every meta-stable state, even if differences between these
states are only minor structural changes. 

Dynamic graphical models (DGM) replaces the global representation
of metastable states with local sub-systems \cite{Olsson_2019}. These
sub-systems are spatially localized and can be a side-chain rotamer
or a whole protein domain. A DGM aims to encode the conformational
states of all the sub-systems and how they influence each other's
evolution in time. Like MSMs, DGMs approximate the transition probability
densities between all possible configurations of our sub-systems,
yet without the need to enumerate them all explicitly. Their indirect
representation of global configurations allows DGMs to rely on fewer
parameters than MSMs, lowering simulation data demands. A recent study
shows that this strategy can be very effective when modeling molecular
dynamics, quantitatively predicting the thermodynamics and kinetics
of molecular systems. As DGMs are generative models, we may also predict
realistic meta-stable states not seen during the model's estimation.

MSMs have come a long way. We now see this methodology's regular application
to the quantitative study of complex problems such as protein folding,
protein-protein interactions, and conformational dynamics. The growing
community of researchers, together with improvements in simulation
methodology and hardware, is rapidly expanding the scope of systems
we can address. With the advent of powerful machine learning-based
we can expect to see these development accelerate further. These developments
and their extensions bode well for Markovian models' future in the
quantitative study of protein-protein encounters.

\section{Acknowledgements}

I thank Dr. Roc\'io Mercado for comments and input on drafts of this
chapter. This work was partially supported by the Wallenberg AI, Autonomous
Systems and Software Program (WASP) funded by the Knut and Alice Wallenberg
Foundation.

\printbibliography

\end{document}